\documentclass[10pt]{iopart}

\usepackage{iopams}  
\usepackage{citesort}
\usepackage{multicol}
\usepackage{lipsum}
\usepackage{cuted}
\usepackage{xcolor}
\usepackage{graphicx,float}
\begin{document}

\title{Generating asymmetric aberration laser beams with controlled intensity distribution}

\author{Sachleen Singh, Vasu Dev and Vishwa Pal$^{*}$}

\address{Department of Physics, Indian Institute of Technology Ropar, Rupnagar 140001, Punjab, India}
\address{{$^*$}Corresponding author: vishwa.pal@iitrpr.ac.in}

\begin{abstract}
We present generation of asymmetric aberration laser beams (aALBs) with controlled intensity distribution, using a diffractive optical element (DOE) involving phase asymmetry. The asymmetry in the phase distribution is introduced by shifting the coordinates in a complex plane. The results show that auto-focusing properties of aALBs remain invariant with respect to the asymmetry parameters. However, a controlled variation in the phase asymmetry allows to control the spatial intensity distribution of aALBs. In an ideal ALB containing equal intensity three bright lobes (for $m=3$), by introducing asymmetry most of the intensity can be transferred to any one of single bright lobe, and forms a high-power density lobe. A precise spatial position of high-power density lobe can be controlled by the asymmetry parameter $\beta$ and $m$, and we have determined the empirical relations for them. We have found that for the specific values of $\beta$, the intensity in the high-power density lobe can be enhanced by $\sim$6 times the intensity in other bright lobes. The experimental results show a good agreement with the numerical simulations. The findings can be suitable for applications such as in optical trapping and manipulation as well as material processing.
\end{abstract}
\ioptwocol

\maketitle
\section{Introduction}
Optical beams with controlled intensity distribution and propagation properties such as robustness against obstructions (self-healing) as well as auto-focusing features are desired for various applications, for example, in imaging, ablation, trapping, guiding the micro-particles, and material processing \cite{zhang:2017,efremidis:2019,dasgupta2011,zhang2021}. Due to their wide applicability, there has been a growing interest in synthesizing such optical beams. Over the years, several types of optical beams with distinct features have been theoretically proposed and realized experimentally. A few examples include Laguerre-Gaussian beam \cite{kim:2015}, Hermite-Gaussian beam \cite{kumar2016}, Bessel-Gauss beam \cite{Acevedo2021}, Airy beam \cite{Wen:2021}, pin-like optical beam \cite{Zhang:19}, abruptly auto-focused beam \cite{zhang:2011}, discrete vortex \cite{Dev:21}, and radial carpet beam \cite{Rasouli:18}.

Recently, a new type of optical beam known as \textit{aberration laser beam} (ALB) has been realized, possessing several unique features \cite{khonina2018b}. Generally, aberrations are considered to be errors in the wavefront of an optical field. Their presence in an optical system results undesired effects such as widening, blurring and distortion in an optical field \cite{born2013principles}. To minimize the effect of aberrations, several efforts have been made \cite{Thurman2019,Cizmar2010,Bowman2010,Booth2002,Love1997}. Further, in designing optical systems with customized properties, aberrations have also been deliberated exploited \cite{Jaffe2021,Dudutis2018,Zolotarev2013,Vera2018,Hernandez2014,Jones1991,Dixit2015}. For example, with the help of a specific type of aberration in an optical system, a brighter focal spot has been obtained in a focal plane \cite{khonina2011,kant2000}. The aberrations have also been explored to characterize the order of singularity in an optical field \cite{Dixit2015,Khonina2004,Serna2001}, creating diffraction-free beams \cite{khonina2018}, and generation of zero intensity spot in the focal plane \cite{Topuzoski2011}. By combining angular dependence of Zernike polynomials with a $r^{q}$ type approximation of chirped Airy function, the ALBs have been realized \cite{khonina2018a,khonina2018b}. It has been shown that for $q>1$, ALBs possess an abrupt auto-focusing. The spectral dependence of auto-focusing of ALBs has also been studied, indicating that auto-focusing distance decreases with the increase in wavelength \cite{Reddy2020}. A study on the robustness of propagation of ALBs in a disorder media has been performed, and several special characteristics such as good self-healing abilities, invariance of auto-focusing distance on disorder strength, controlled variation of auto-focusing distance from a small to large values, have been observed \cite{Dev2021}. The auto-focusing has also been investigated in multiple variants  such as radially polarised circular Airy beam \cite{Liu2013}, vortex circular Airy beam(CAB) \cite{Davis2012,Jiang2012}, circular arrayed CAB \cite{Lai2019}, chirped CAB \cite{Zhang2017}, tightly focused CAB \cite{Chen2017,Zhuang2020}, partially coherent CAB \cite{Jiang2018} and ring Airy Gaussian beam (RAiG) \cite{Hong2020}. As mentioned ALBs possess several special characteristics, but so far, a controlled transfer of intensity from one region to another in ALBs have not been explored. However, it is strongly desired for several applications such as in imaging, ablation, trapping, guiding the micro-particles, and material processing. In the present work, we have investigated the controlled intensity distribution of ALBs by exploiting asymmetry in the phase distribution. 

It is known that with a complex source point the spherical wave solutions can represent a fundamental Gaussian beam with $k$-vectors close to $z$-axis \cite{Deschamps1971,zauderer1986}. Using complex-valued shift of the beam's complex amplitude in Cartesian coordinates, this idea has been extended to known solutions of paraxial wave equation \cite{Kotlyar2020a,Kotlyar2017,Kotlyar2014,Kotlyar2014a,Kovalev2016}. As a result of it crescent shape appears in the symmetric beam structure such that a part of beam becomes brighter than the rest \cite{Kovalev2016}. These so called asymmetric beams also comprise of fractional orbital angular momentum, which further depends upon asymmetric parameters \cite{Kovalev2016}. By varying these parameters, the beam profile in transverse plane gets modified and for a certain set of values, the regular beam profile reappears. Thus asymmetric beams can be also be considered as a generalized version of their symmetric counterparts. These beams  have  been used to generate  a pair of entangled photons with broad orbital angular momentum using spontaneous parametric down-conversion \cite{Alam2018}, and also for designing optical tweezers for micro-manipulation of  small sized particles \cite{Kotlyar2016,Kovalev2016a}.

So far, asymmetric effects have been studied mostly with the beams possessing intrinsic symmetry. In the present work, we have investigated asymmetric effects in the aberration laser beams, for controlling their intensity distribution. A detailed analysis of asymmetric effects with quantification is presented. In Sec.\,\ref{II}, we have given an analytical description of asymmetric aberration laser beams (aALBs). In Sec.\,\ref{III}, we described experimental generation and numerical simulations of aALBs. The propagation properties of aALBs are compared with an ideal ALB. In Sec.\,\ref{IV}, we described the role of different asymmetric parameters on intensity distribution of aALBs. In Sec.\,\ref{V}, we present results on spatial control of intensity distribution in aALBs. Finally, in Sec.\,\ref{VII} concluding remarks are presented. 
\section{Theoretical description}\label{II}
 The expression of phase of ALB is given by \cite{Reddy2020}
\begin{equation}  
\zeta(r,\phi) = \exp(-i\alpha r^q + i\sin(m\phi)) \quad r \leq R \label{eq1}
\end{equation}
such that
\begin{equation}
r\exp(i\phi) = x + iy,~~~  r^2 = x^2 + y^2. \label{eq2}
\end{equation}
Here, $q$ is a positive number that represents an arbitrary radial power, $m$ is an integer that denotes periodic angular dependence, and $\alpha$ is a scale parameter, has a dimension of $\mathrm{mm}^{-q}$, $R$ is the radius of the diffractive optical element (DOE).

The term $\exp(-i\alpha r^q)$ represents the transmission function of a generalized parabolic lens \cite{Ustinov2013}. An optical element encoded with the above phase function is same  as that of a  zone plate with circular lines. The separation between adjacent circles increases linearly when $q=2$. The diffractive version of such a phase element will be equivalent to a classical lens (quadratic phase dependence) \cite{khonina2018a}. As pointed by earlier studies \cite{Reddy2020,Dev2021}, the ALB shows more abrupt focusing for $q=2$, therefore in the present work the numerical simulations and experiment are performed for $q=2$. However, for other values of $q$ a general expression of aALB has been provided in \ref{appendixA}. Using the complex coordinate shifting in Eq.\,(\ref{eq1}) \cite{Kovalev2016}, we have
\begin{equation}
\zeta = \exp(-i\alpha s^2 + i \sin( m\theta)) \label{eq3}, 
\end{equation}
where,
\begin{eqnarray}
s^2 = (x - x_o)^2 + (y - y_o)^2, \label{eq4} \\
x_o = a + ib, \quad y_o = c + id,  \quad \, a,b,c,d \in\mathbb{R}. \label{eq5}
\end{eqnarray}
For the new polar coordinates $p$ and $\theta$: 
\begin{eqnarray}
p~\exp(i\theta) &= (x - x_o) + i(y - y_o) \nonumber\\ 
         &= (x - a + d )+ i(y - c -b). \label{eq6}
\end{eqnarray}
 The angular coordinate $\theta $ is given by
\begin{eqnarray}
p~\cos(\theta) = (x - a +d), ~ p~\sin(\theta)= y - c- b, \label{eq7}\\
\theta = \tan^{-1}\left(\frac{y - c - b}{x - a +d}\right). \label{eq8}
\end{eqnarray}
The modified phase due to complex coordinate shifting can be calculated from Eq.\,(\ref{eq3}). For simplification, we have taken $x-a=X$ and $y-c=Y$. Thus, 
\begin{eqnarray} 
 s^{2}&=&(X^2 + Y^2 - b^2 - d^2) - i(2Xb + 2Yd), \label{eq9}\\
 &=&G~\exp(i\gamma), \label{eq10}
\end{eqnarray}
where,
\begin{eqnarray}\label{4}
G~\cos(\gamma) &=& (X^2 + Y^2 - b^2 - d^2), \label{eq11}\\
G~\sin(\gamma) &=& -2(Xb + Yd).\label{eq12}
\end{eqnarray}
Substituting $s^2$ into Eq.\,(\ref{eq3}), we get  
\begin{eqnarray}
\zeta =\exp(-i\alpha G\cos(\gamma)+i\sin(m\theta))\exp(\alpha G\sin(\gamma)).\label{eq13}
\end{eqnarray}
The term $\exp(\alpha G\sin(\gamma))$ is a real quantity and does not contribute to the phase, thus the phase expression is reduced as
\begin{eqnarray}
\zeta =\exp(-i\alpha G\cos(\gamma)+i\sin(m\theta)),\nonumber\\
~~=\exp(-i\alpha\,((x-a)^2 + (y-c)^2 - b^2 - d^2) \nonumber \\ 
~~~~~~+ i\sin(m\theta)).\label{eq14}
\end{eqnarray}
The expression for $\theta$ and its dependence on parameters is given by Eq.\,(\ref{eq8}). It can be readily noticed that the parameters $a$ and $c$ (real part of $x_0$ and $y_0$) are just shifting the origin in a real two dimensional (2D) plane. From a practical point of view, it will just simulate the effect for misaligned or off-axis beam. By choosing $a = c = 0$, we impose axis-to-axis alignment of an input beam with the DOE. Thus new modified phase for ALB can be expressed as
\begin{equation}
\zeta =\exp(-i\alpha~(x^2 + y^2 -(b^2 +d^2)) + i \sin(m\theta)), \label{eq15}
\end{equation}
where,
\begin{equation}
\theta =  \tan^{-1}\left(\frac{y - b}{x - (-d)}\right). \label{eq16}
\end{equation}
It can be seen that for the part of chirped phase asymmetry parameters $b$ and $d$ are just rescaling the radius of concentric circles (origin ($0$, $0$)). Whereas, in trigonometric term ($\exp(i\sin(m\theta)$), asymmetry shifts the origin to ($b,-d$) locally. The asymmetry parameters $b$ and $d$ can be expressed in terms of angular parameters $(w, \beta)$, which only relocates the origin for trigonometric phase without changing the functional form of chirped phase, as described below.
\begin{equation}
d = -w\,cos(\beta), \quad b = w\,\sin(\beta). \label{eq17}
\end{equation}
Considering an input laser beam with fundamental transverse mode (Gaussian), the optical field of an asymmetric ALB (aALB) can be written as 
\begin{equation}
\eqalign{U(r,\theta)= \exp\left(\frac{-r^2}{2\sigma^2}\right)\exp\left(-i\alpha(r^2 -w^2) + i\sin(m\theta)\right)}. \label{eq18}
\end{equation} 
The condition $ 0 < w^2 < 2\sigma^2$ makes sure that the asymmetric effects are well aligned with an input Gaussian beam. It should be noted that the symmetry of trigonometrically modulated phase is decided by parity of $m$ \cite{Khonina2012}. As $\sin(m(\theta + \pi)) = \pm \sin(m\theta)$, the diametrically opposite points on phase distribution will be in and out of phase for even and odd parity, respectively. In earlier studies it has been shown that the diffraction pattern of such a phase distribution consists of a number of bright spots depending upon the value of $m$ \cite{Topuzoski2011}. In our work, we have considered the case of $m=3$, however, similar findings can be obtained for other values of $m$. For $\sigma =1.45$ mm and $m=3$, we find that the value of $w = 1$ mm introduces appreciable asymmetry into the phase of aALB. So, we have kept these parameters same throughout the work. 

\begin{figure}[htbp]
\centering
\includegraphics[height = 6cm,width = 8cm,keepaspectratio= true]{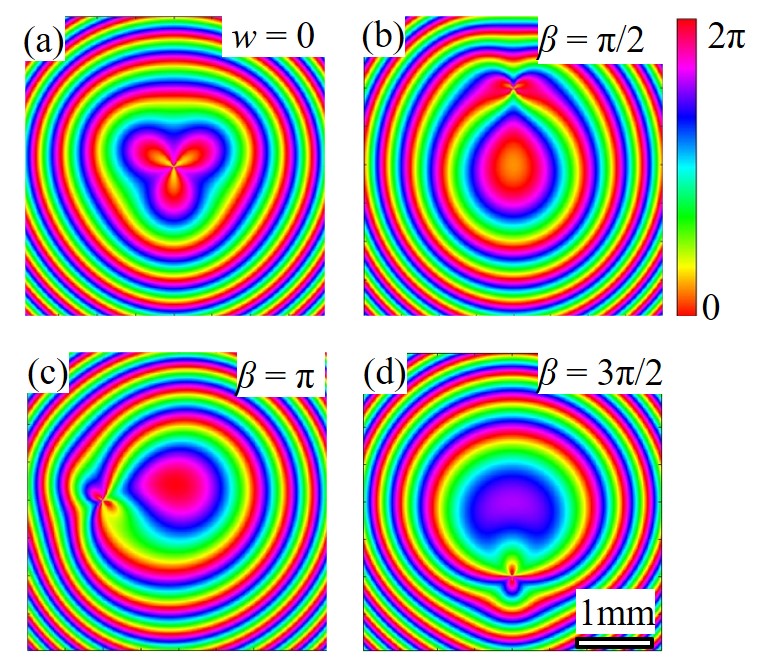}
\caption{The phase distribution of DOE with (a) $w = 0,~\beta = [0, 2\pi]$, (b) $w = 1$ mm, $\beta = \pi/2$, (c) $w = 1$ mm, $\beta = \pi$, (d) $w = 1$ mm, $\beta = 3\pi/2$. The other parameters are $m = 3$, $q = 2$, $\alpha=5.9$ mm$^{-2}$.}
\label{fig1}
\end{figure}

The phase distributions of DOEs for generating the ideal and asymmetric ALBs are shown in Fig.\,\ref{fig1}. Figure\,\ref{fig1}(a) shows the DOE of an ideal ALB ($w = 0$). As substituting $w=0$ in Eqs.\,(\ref{eq15})-(\ref{eq17}), gives rise to Eq.\,(\ref{eq1}) for an ideal ALB. Figures\,(\ref{fig1}(b)-\ref{fig1}(d)) show the DOEs of an asymmetric ALB (aALB) with different asymmetry parameter $\beta$. As evident, different values of $\beta$ provides a different phase distribution of DOE, and thereby provides an ability to control the propagation properties of aALBs.
\section{Propagation of aALBs}\label{III}
\begin{figure*}[htbp]
\centering
\includegraphics[width = 15cm,keepaspectratio = true]{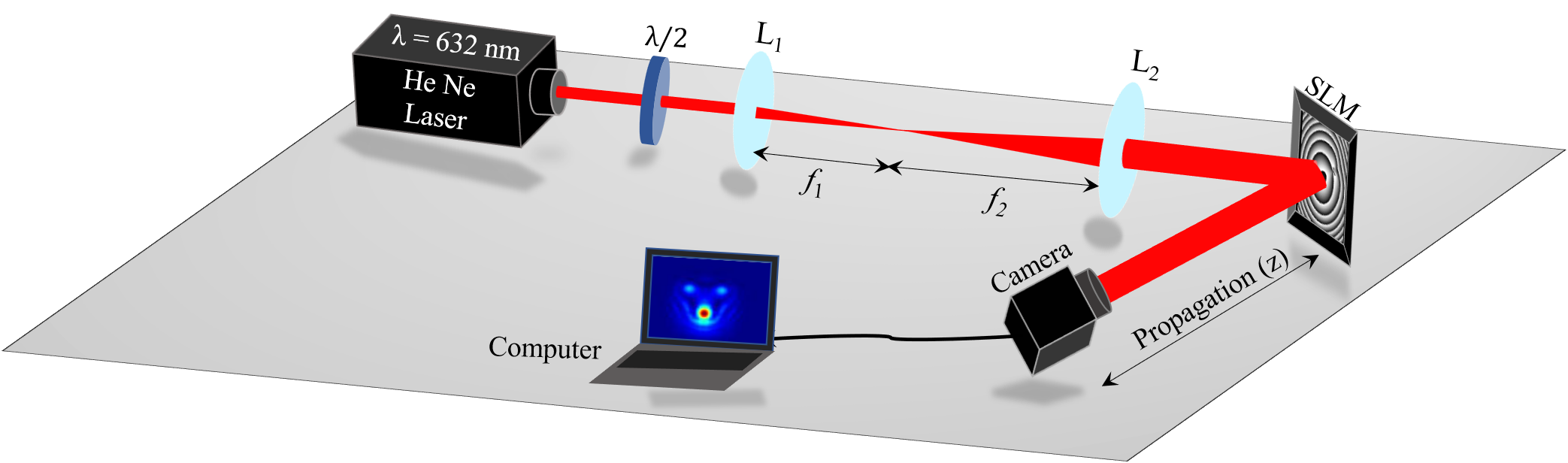}
\caption{Experimental arrangement for generation and characterization of asymmetric ALB. L$_{1}$ and L$_{2}$: plano-convex lenses of focal length $f_{1}=5$ cm and $f_{2}=20$ cm, respectively. $\lambda/2$: half-wave plate, SLM: spatial light modulator.}
\label{fig2}
\end{figure*}

The experimental arrangement for the generation and characterization of aALB is schematically shown in Fig.\,\ref{fig2}. A linearly polarized light from a He-Ne laser ($\lambda=632$ nm) incidents on a half-wave ($\lambda/2$) plate to fix the polarization orientation in a specific direction. After that light is passed through a telescope made with lenses L$_{1}$ ($f_{1}=5$ cm) and L$_{2}$ ($f_{2}=20$ cm) to magnify it's size in order to illuminate well the screen of a phase-only spatial light modulator (SLM). We impose a phase pattern (DOE) on the SLM, which modulates the phase of an incident light, and after propagating a certain distance aALB is formed. The intensity distribution of aALB at different propagation distances is recorded on a CCD camera.

Further, we have numerically simulated the propagation of an aALB using an extended Huygens-Fresenel integral as
\begin{eqnarray}
U(\rho,\phi,z) = -\frac{ik_o}{2\pi z}\exp(ik_oz)exp\left( \frac{ik_o}{2z}\rho^2\right) \nonumber \\ 
                \times\int\int U(r,\theta)\exp\left( \frac{ik_o}{2z}r^2\right)\exp\left(-\frac{ik_o}{z}\rho r \cos(\theta - \phi)\right) \nonumber\\
                ~~~~~~~\times rdr d\theta, \label{eq19}
\end{eqnarray}
where ($r,\theta$) and ($\rho, \phi$) represent coordinates of source and observation (output) planes, respectively, separated by a distance $z$. $k_o = 2\pi/\lambda$ represents the wave number of an optical field in free space. 

\begin{figure*}[htbp]
\centering
\includegraphics[width = 12cm,keepaspectratio = true]{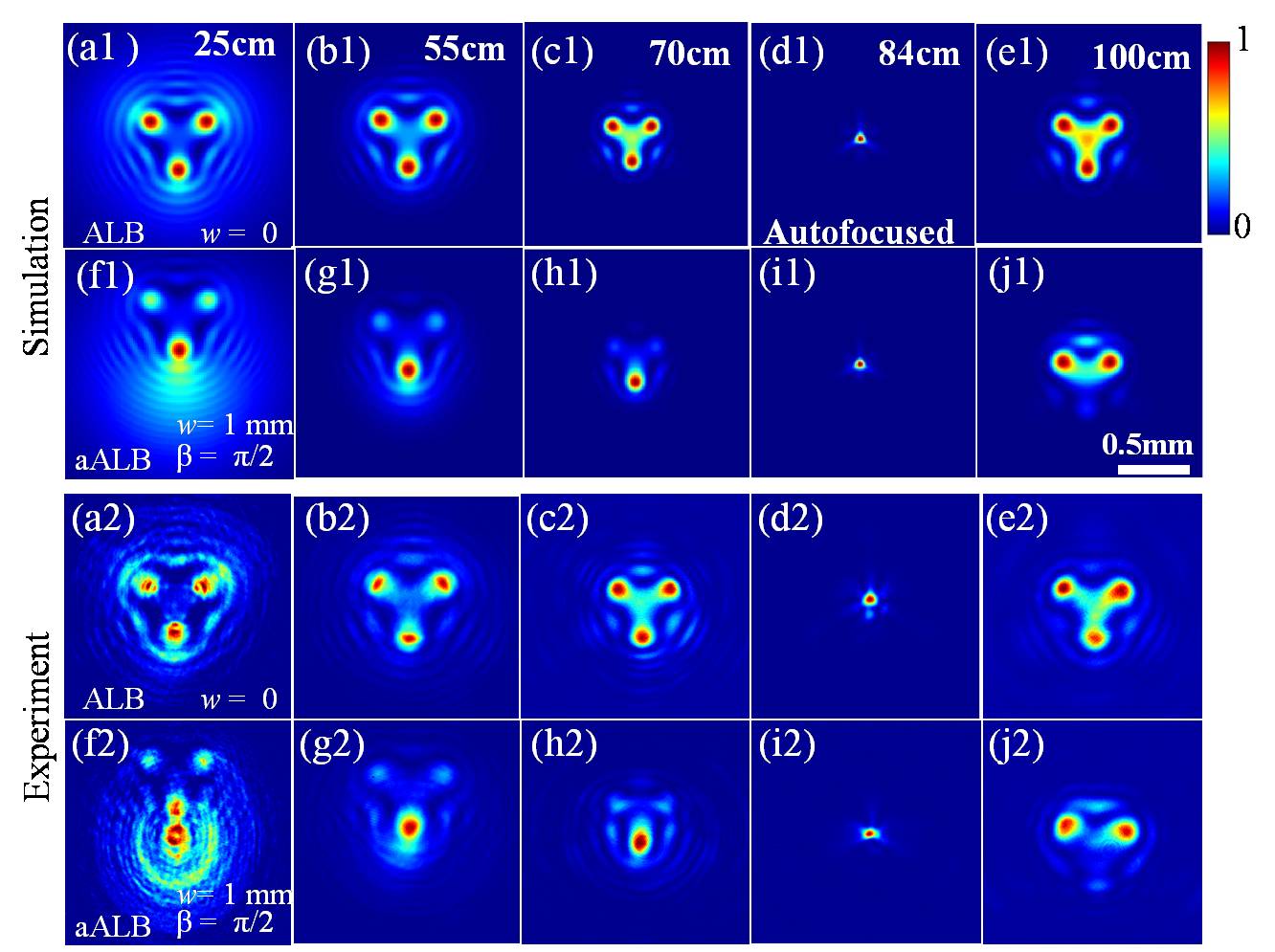}
\caption{The intensity distributions of ideal ALB ($w=0$) ((a1)-(e1), (a2)-(e2)) and aALB ($w=1$ mm, $\beta=\pi/2$) ((f1)-(j1), (f2)-(j2)) at different propagation distances $z=25$ cm, $55$ cm, $70$ cm, $84$ cm, and $100$ cm. The results are obtained for the following parameters: $\alpha = 5.9$\,mm$^{-2}$, $\sigma = 1.45$\,mm, m = 3, q = 2\; and\; $ \lambda = 632 $ \,nm.} 
\label{fig3}
\end{figure*}

The experimental and simulation results are shown in Fig.\,\ref{fig3}, presenting a comparison between the propagation of ideal ALB and aALB. The results are presented for parameter values of $m=3$, $q=2$, $\alpha=5.9$ mm$^{-2}$. For a aALB, the asymmetry parameters are taken as $w=1$ mm and $\beta=\pi/2$.
Figures.\,\ref{fig3}(a1)-\ref{fig3}(e1)) and Figs.\,\ref{fig3}(a2)-\ref{fig3}(e2) show the simulation and experimental results
of intensity distributions of an ideal ALB at different propagation distances, respectively. As evident, for $m=3$ there are
three bright lobes having equal intensity within them. During the propagation, the intensity from background moves inside the
bright lobes, and evolution of intensity remains symmetrically as all three bright lobes are equally intense
(Figs.\,\ref{fig3}(a1)-\ref{fig3}(c1), Figs.\,\ref{fig3}(a2)-\ref{fig3}(c2)). At a distance $z=70$ cm, the bright lobe pattern
is fully developed, as most of the intensity from all parts of the beam shifted equally into them. After further propagation,
the bright lobe pattern starts shrinking by merging of bright lobes. At a distance $z=84$ cm, all three bright lobes collapsed
into a tightly focused single bright spot, called the auto-focusing distance ($z = z_{af}$)
(Figs.\,\ref{fig3}(d1) and \ref{fig3}(d2)). After auto-focusing distance, the bright lobes again get separated and intensity
distributes symmetrically and equally among them. Figures.\,\ref{fig3}(f1)-\ref{fig3}(j1)) and
Figs.\,\ref{fig3}(f2)-\ref{fig3}(j2) show the simulation and experimental results of intensity distributions of aALB at
different propagation distances, respectively. As evident, by introducing asymmetry using complex coordinate shifting ($w=1$
mm and $\beta=\pi/2$), the symmetry of equal intensity distribution in three bright lobes gets disturbed
(Figs.\,\ref{fig3}(f1)-\ref{fig3}(h1) and Figs.\,\ref{fig3}(f2)-\ref{fig3}(h2)). More specifically, instead of three equal
intensity bright lobes, the intensity in the bottom single bright lobe is larger than the top two bright lobes. Further, as
aALB propagates there is a continuous transfer of intensity between the bright lobes. At $z=70$ cm, the intensity in the upper
two bright lobes reduced dramatically, and enhanced significantly in the bottom bright lobe (Figs.\,\ref{fig3}(h1) and
\ref{fig3}(h2)). It should be noticed that, for an ideal ALB, the lobes pattern develops around on-axis (beam axis)
(Fig.\,\ref{fig3}(a1), \ref{fig3}(a2)). Whereas, for aALB, initially the lobes pattern develops around the point ($w$,
$\beta$), and after that it moves towards on-axis. At auto-focusing distance $z=84$ cm, again a tightly focused single bright
spot centered on the beam axis is observed (Figs.\,\ref{fig3}(i1) and \ref{fig3}(i2)). After auto-focusing distance, the
intensity again redistributes such that it becomes larger in the upper two bright lobes as compared to the bottom bright lobe
(Figs.\,\ref{fig3}(j1) and \ref{fig3}(j2)).

We have found that the auto-focusing distance remains the same for both ideal ALB and aALB. However, for aALB, the asymmetry leads to a continuous variation of intensity distribution with the propagation distance. The asymmetry induced variation in the intensity distribution has also been observed with other beams such as LG, Bessel, Bessel-Gauss and Kummer laser beams \cite{Kotlyar2020a,Kotlyar2017,Kotlyar2014,Kotlyar2014a,Kovalev2016}. For an ideal ALB, the relationship between the beam parameters and auto-focusing distance is given as \cite{Reddy2020}
\begin{equation} 
z_{af} \approx \frac{2\pi}{q \alpha \lambda (2\sigma /3)^{q-2}}.\label{eq20}
\end{equation}
By tuning the beam parameters, the auto-focusing distance can be varied from the small to large values.

\begin{figure*}[htbp]
\centering
\includegraphics[width = 14.0cm]{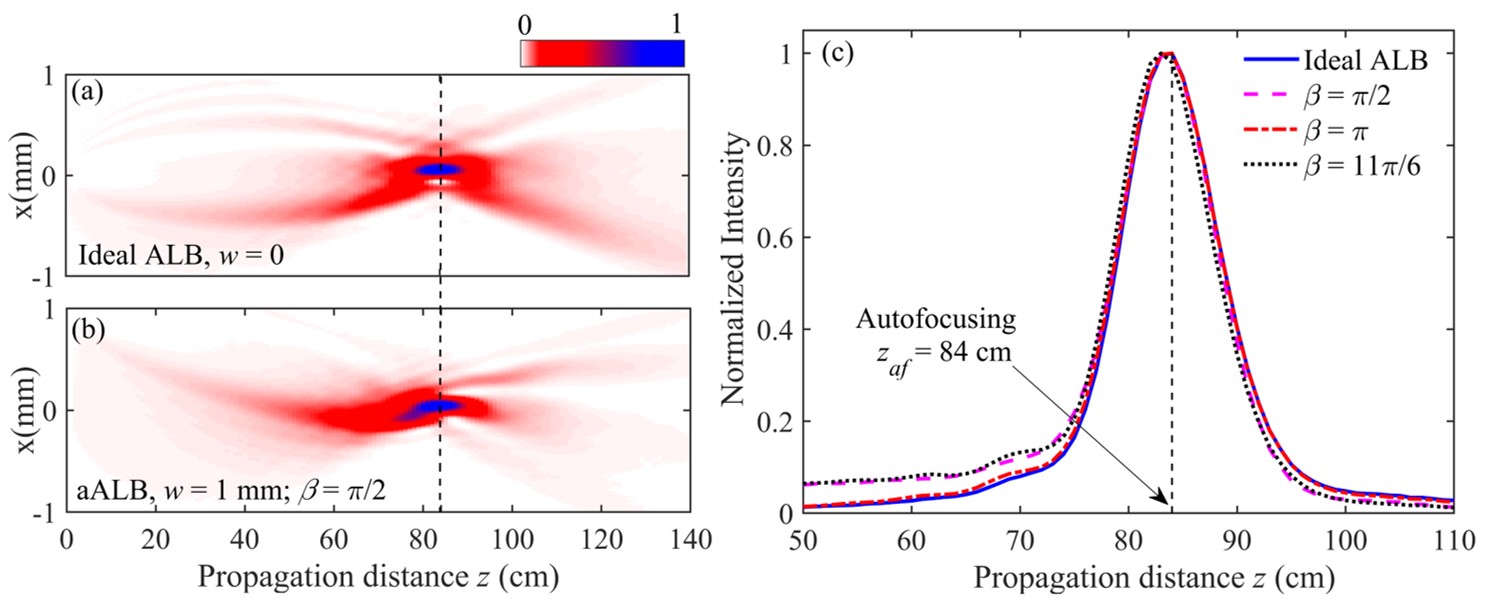}
\caption{(a) Intensity distribution of an ideal ALB as a function of propagation distance. (b) Intensity distribution of an aALB (with $w=1$ mm and $\beta=\pi/2$) as a function of propagation distance. (c) Longitudinal intensity cross-section taken along the horizontal axis (at $x=0$) (on-axis intensity) in (a) and (b), as a function of $z$. The blue solid curve: ideal ALB; dashed pink curve: aALB with $\beta = \pi/2$; dash-dotted red curve: aALB with $\beta = \pi$; dotted black curve: aALB with $\beta = 11\pi/6$. The parameters are taken as $w=1$ mm, $\sigma = 1.45$ mm, $\alpha = 5.9$ mm$^{-2}$, $m =3$, $q = 2$ and $\lambda=632$ nm. $z_{af}$ denotes the auto-focusing distance.}
\label{fig4}
\end{figure*}

To check the dependence of auto-focusing distance on asymmetry parameter $\beta$, we have propagated aALB with different $\beta$ values, and then analyzed on-axis intensity distribution. The results are shown in Fig.\,\ref{fig4}. Figures\,\ref{fig4}(a) and \ref{fig4}(b) show the propagation of an ideal ALB and aALB with $\beta=\pi/2$. As evidenced, the on-axis intensity peak is observed at $z=84$ cm in both cases (marked by a vertical dashed-line), denotes the auto-focusing distance. Figure\,\ref{fig4}(c) shows the longitudinal intensity cross-section taken along the horizontal axis (at $x=0$), indicating that on-axis intensity peak occurs at the same $z$ value for ideal ALB and aALB with different $\beta$ values, which shows that the auto-focusing distance remains invariant with respect to the asymmetry. Note, the 2D intensity plots for aALB with $\beta=\pi$ and $11\pi/6$ are not shown. The observed value of auto-focusing distance agrees with the calculated value of $z_{af}=84.25$ cm (Eq.\,(\ref{eq20})) for an ideal ALB.
\section{Intensity Distribution for different $\beta$}\label{IV}
\begin{figure*}[htbp]
\centering
\includegraphics[width = 13.8cm,keepaspectratio = true]{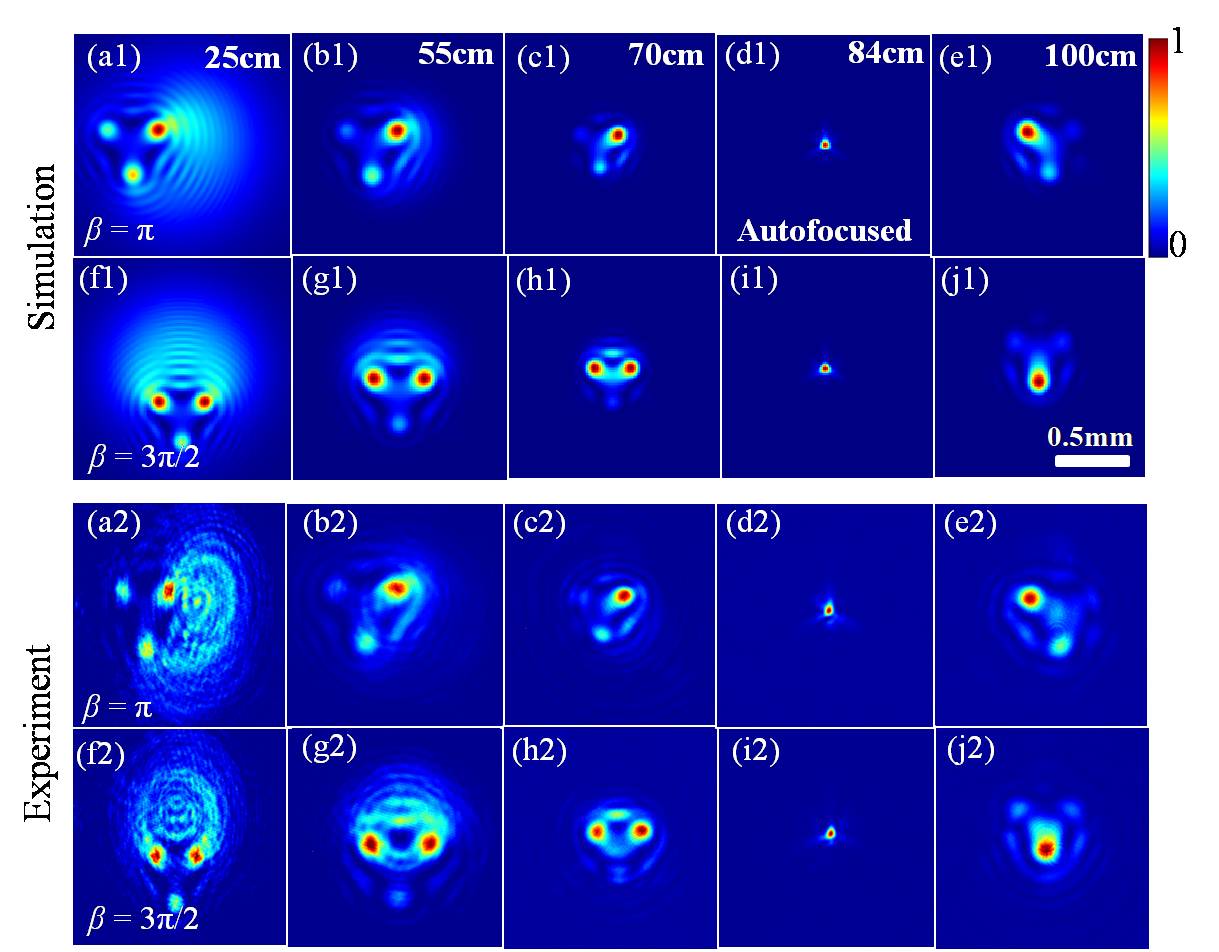}
\caption{Intensity distribution of aALB at various propagation distances, for different values of $\beta$. ((a1) - (e1), (a2) - (e2)) $\beta = \pi/2$, and ((f1)-(j1), (f2) - (j2)) $\beta = 3\pi/2$. The results are obtained for the following parameters: $\alpha = 5.9$\,mm$^{-2}$, $\sigma = 1.45$\,mm, $m = 3$, $q = 2$\; and\; $\lambda = 632 $ \,nm.}
\label{fig5}
\end{figure*}
To generalize and gain better understating, we have further varied the phase asymmetry, and analyzed it's effect on the intensity distribution of aALBs. The results are shown in Fig.\,\ref{fig5}. Figures.\,\ref{fig5}((a1)-(e1)) and Figs.\,\ref{fig5}((a2)-(e2)) show the simulation and experimental results for an asymmetry parameter $\beta=\pi$, respectively. Figures\,\ref{fig5}((f1)-(j1)) and Figs.\,\ref{fig5}((f2)-(j2)) show the results for $\beta=3\pi/2$. Note, for $\beta=\pi/2$, the results are shown in Fig.\,\ref{fig3}. As evident, the asymmetry leads to a significant change in the intensity distribution of aALB. For different values of $\beta$ the intensity in the background as well as inside bright lobes migrates differently, and thereby enabling a controlled intensity distribution of aALB. For example, for $\beta=\pi$ a greater portion of intensity can be transferred to a single bright lobe (Figs.\,\ref{fig5}(c1) and \ref{fig5}(c2)). Whereas, for $\beta=3\pi/2$, the intensity transfers into two bright lobes (Figs.\,\ref{fig5}(h1) and \ref{fig5}(h2)). By a careful choice of $\beta$ values, one can control the intensity distribution precisely. As mentioned above, for any asymmetry value, initially (for small $z$ values) beam develops around the coordinates ($w$, $\beta$), so spatial position of three bright lobes pattern (triangular pattern for $m=3$) appears off-centered (Figs.\,\ref{fig5}(a1), \ref{fig5}(f1), \ref{fig5}(a2) and \ref{fig5}(f2)) from the on-axis. As the beam propagates for longer $z$ values, in addition to migration of intensity into bright lobes, the spatial position of bright lobes pattern also moves towards the on-axis center. At auto-focusing distance $z=84$ cm, the intensity becomes tightly focused to a single bright spot (Figs.\,\ref{fig5}(d1), \ref{fig5}(i1), \ref{fig5}(d2) and \ref{fig5}(i2)). After the auto-focusing distance (at $z=100$ cm), the intensity distribution again changes depending on the value of $\beta$.
\begin{figure}[htbp]
\centering
\includegraphics[height = 6.4cm,keepaspectratio = true]{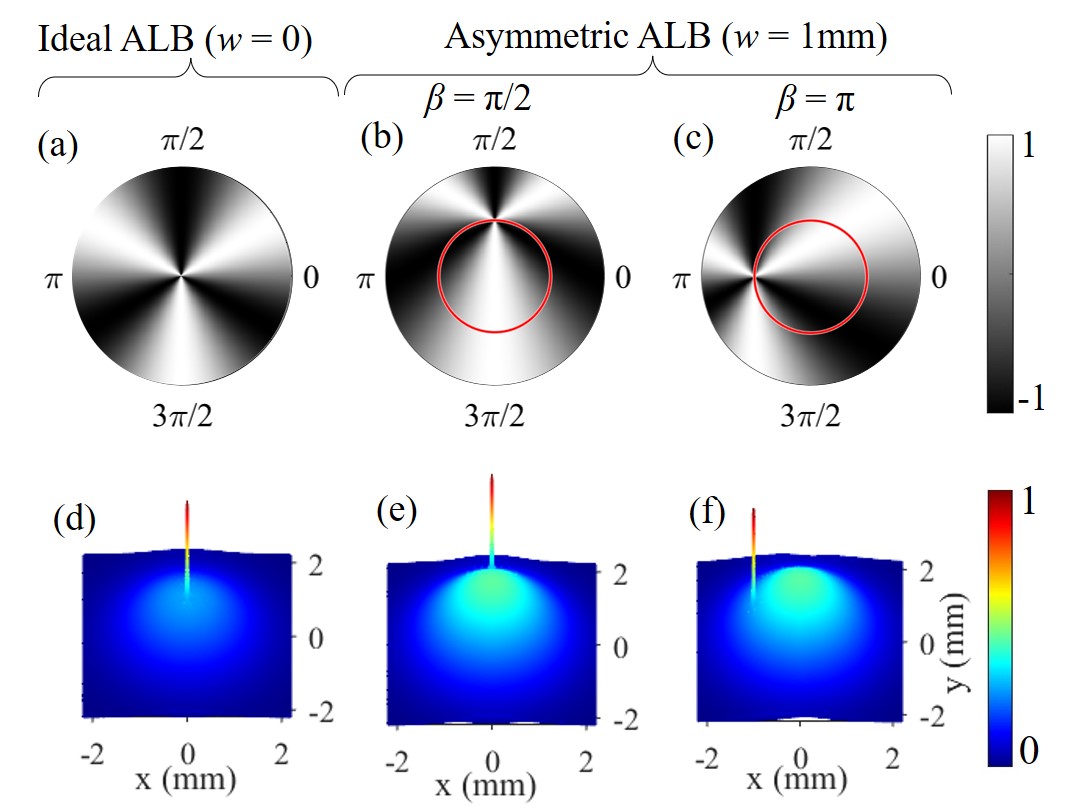}
\caption{Trigonometric Phase (exp($i\, \sin(3\theta)$)) for (a) $w = 0$ and $\beta = [0, 2\pi]$, (b) $w = 1$ mm, $\beta = \pi/2$, (c) $w = 1$ mm, $\beta = \pi$. A red circle of radius 1 mm with centre(0,0) is drawn to show the effect of asymmetry. (d-f) The intensity distribution of ideal ALB and aALB at $z = 0.027$ cm (near-field), corresponding to the phase distributions given in (a)-(c). The other simulation parameters are taken as $\alpha=5.9$ mm$^{-2}$, $\sigma=1.45$ mm, $m=3$, $q=2$, and $\lambda=632$ nm.}
\label{fig6}
\end{figure}

The off-centered position of bright lobes pattern is quite evident at distance $z=25$ cm (Figs.\,\ref{fig5}(a1), \ref{fig5}(f1), \ref{fig5}(a2) and \ref{fig5}(f2)) as well as for the small values of $z$ (\ref{appendixC}). This can be attributed to the fact that the spatial position of indeterminate phase point \cite{Henault2016} of trigonometric phase ($\exp(i\sin(3\theta)))$ is correlated with the asymmetry parameters ($w,~\beta$) (Eqs.\,(\ref{eq15})-(\ref{eq17})), as shown in Fig.\,\ref{fig6}. For an ideal ALB, the on-axis center of an input beam and indeterminate phase point coincides, and as a result of propagation it leads to the formation of a three bright lobes pattern with triangular symmetry (Figs.\,\ref{fig6}(a)). However, for aALB (non-zero value of $w$), the indeterminate phase point does not coincides with the on-axis center of an input beam. Figures\,\ref{fig6}(b-c) show the change in the position of indeterminate phase point with the coordinates ($w,~\beta$). A red circle of radius $1$ mm with center ($0$,$0$) is drawn to show the shifting of indeterminate phase point from the center. This shift due to the asymmetry affects significantly the intensity distribution of aALB during the propagation. When both ideal ALB and aALB are propagated by a small distance $z=0.027$ cm (near-field plane (close to DOE plane)), a peak with maximum intensity (say maximum intensity point (MIP)) on top of Gaussian distribution starts appearing at the precise location of indeterminate phase point (Figs.\,\ref{fig6}(d-f)), which then helps to initially develop the bright lobes pattern in aALB, off-centered at coordinates ($w$, $\beta$). Upon propagation to large distances, this leads to an asymmetric intensity distribution, which can be precisely controlled by varying $w$ and $\beta$, as shown in Figs.\,\ref{fig3} and \ref{fig5}.

\begin{figure}[htbp]
\centering
\includegraphics[height = 5cm, keepaspectratio = true]{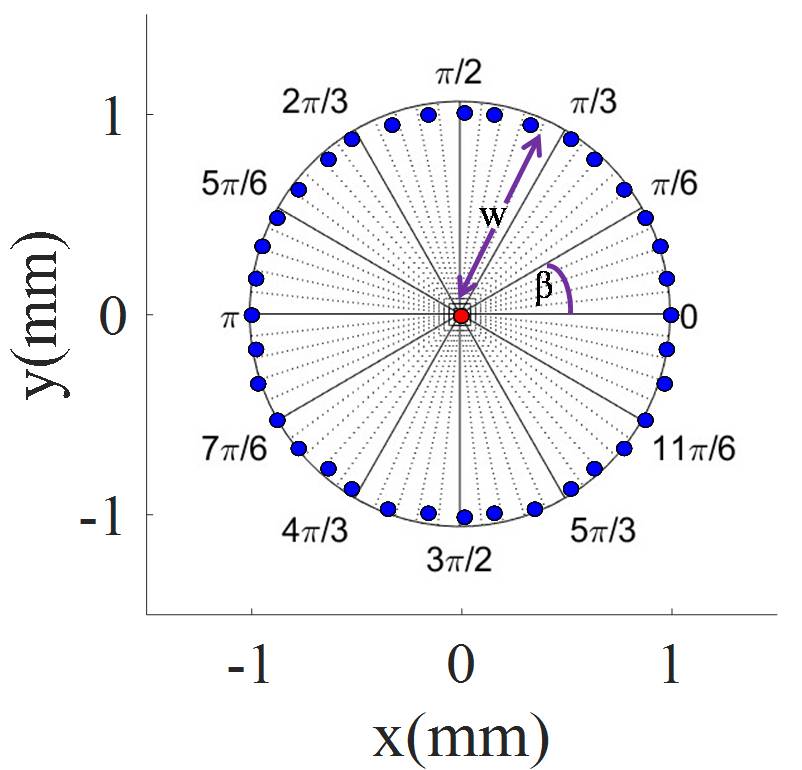}
\caption{The positions of near-field MIP for different asymmetry parameters $w$ and $\beta$. The position of near-field MIP is extracted from aALB at $z=0.027$ cm. Red filled circle: MIP for an ideal ALB ($w=0$); Blue filled circle: MIP for aALB ($w=1$ mm).}
\label{fig7}
\end{figure}

A detailed plot for varying the position of near-field MIP with asymmetry parameters ($w$, $\beta$) is shown in Fig.\,\ref{fig7}. 
The position of near-field MIP is extracted from aALB at $z=0.027$ cm (Figs.\,\ref{fig6}(d-f)). In Fig.\,\ref{fig7}, a red filled circle in the center represents the location of a near-field MIP for ideal ALB (a special case of aALB for $w=0$), and blue filled circles denote the near-field MIPs for aALB ($w=1$ mm).
\section{Spatial control of intensity distribution}\label{V}
For various applications, we require lobes with high power density with controlled position in spatial domain. The question is that for which set of asymmetry parameters such high power density lobes can be obtained at different spatial positions. So we have explored the correlation between the asymmetry parameters and the spatial position of bright lobes.
 
From Fig.\ref{fig4}(c), it can be seen that a sharp rise of intensity due to auto-focusing phenomenon occurs near the value of $z=70$ cm. After this distance, the lobes in the pattern begins to merge and converts into a tightly focused bright spot at $z_{af}=84$ cm. At the auto-focusing distance, the peak power becomes maximum due to merging of bright lobes, but it always appears at on-axis center of the beam. Thus spatial position of the tightly focused bright spot at auto-focusing distance can not be changed. However, upto a distance of $z = 70$ cm the high intensity lobes continues to develop and remain separated well in a pattern. At $z=70$ cm, most of the intensity becomes confined tightly within these bright lobes (lobes with high-power density) (Figs.\,\ref{fig3} and \ref{fig5}). At this distance, the spatial position of high-power density lobe can be controlled by transferring power between the bright lobes using asymmetry parameter $\beta$. Note, a similar control can also be obtained at other distance after auto-focusing, as these bright lobes are separated well and intensity is confined within them.

For different asymmetry parameters, the intensity distributions of aALB are given in Figs.\,\ref{fig3}, \ref{fig5} and \ref{appendixB}. As evident, the spatial intensity distribution of aALB is different for different values of $w$ and $\beta$. For the specific values of $\beta$, most of the intensity can be confined to any one of the bright lobes. For example, for $\beta = \pi/2$ and $w=1$ mm, most of the intensity shifts and confines to a single bright lobe, which represents a high-power density lobe (Figs.\,\ref{fig3}(h1) and \ref{fig3}(h2)). The spatial position of high-power density lobe can be varied by choosing other specific values of $\beta$, for which most of the intensity will transfer to other bright lobe.

\begin{figure}[htbp]
\centering
\includegraphics[height = 7.5cm, keepaspectratio = true]{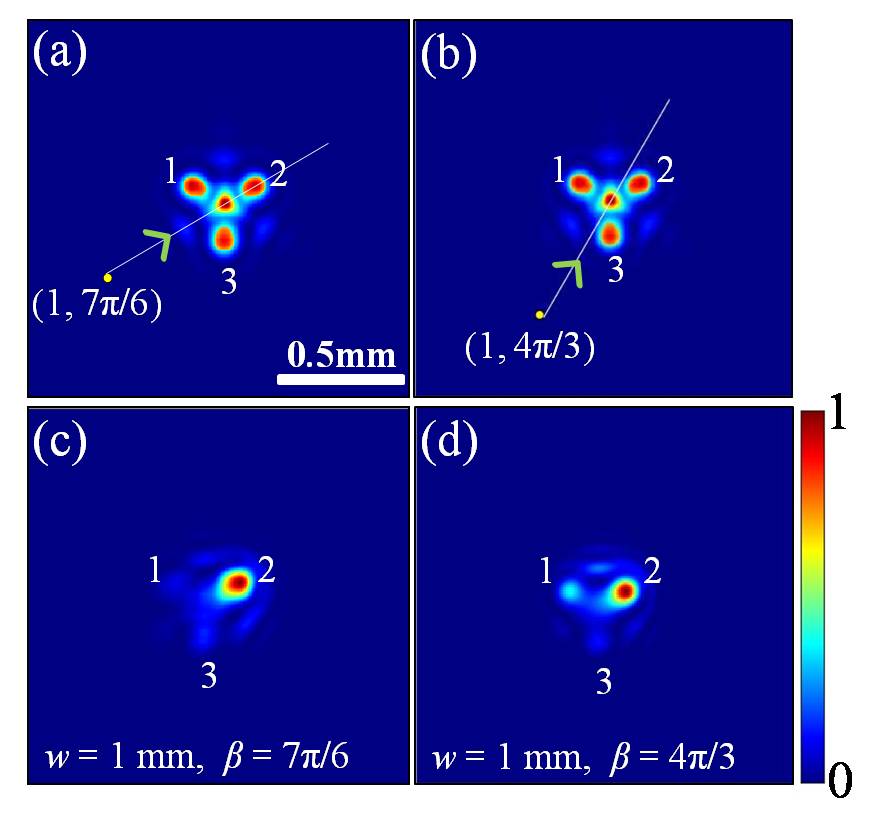}
\caption{(a)-(b) The intensity distributions of ideal ALB at $z=70$ cm (Fig.\,\ref{fig3}(c1)) and at auto-focusing distance $z_{af}=84$ cm (Fig.\,\ref{fig3}(d1)) are superimposed. The yellow dots in (a) and (b) mark the spatial position of near-field MIP (Fig.\,\ref{fig7}) for $w=1$ mm and $\beta= 7\pi$/6 and $4\pi$/3, respectively. (c)-(d) The intensity distributions of aALB at $z=70$ cm for $w=1$ mm and $\beta$=$7\pi/6$ and $4\pi/3$. Other parameters are taken as $m=3$, $q=2$, $\alpha=5.9$ mm$^{-2}$, $\sigma=1.45$ mm and $\lambda=632$ nm.}
\label{fig8}
\end{figure}

A controlled shift of intensity into any one of bright bright lobes can be explained by establishing a correlation between the auto-focusing point ($z=z_{af}$) (Fig.\,\ref{fig3}(d1)), near-field MIP (Fig.\,\ref{fig7}) and position of bright lobes in the pattern (Fig.\,\ref{fig3}(c1)). Particularly, a relative alignment among them decides the shifting of intensity in a certain direction, and accordingly high-power density lobe forms at a specific spatial position. For two set of parameters ($w=1$ mm, $\beta=7\pi/6$) and ($w=1$ mm, $\beta=4\pi/3$), the results are shown in Fig.\,\ref{fig8}. 

In Figs.\,\ref{fig8}((a)-(b)), the intensity distributions of ideal ALB at $z=70$ cm (Fig.\,\ref{fig3}(c1)) and at auto-focusing distance $z_{af}=84$ cm (Fig.\,\ref{fig3}(d1)), are superimposed. The spatial position of MIPs (Fig.\,\ref{fig7}) for two set of asymmetry parameters ($w=1$, $\beta=7\pi/6$) and ($w=1$, $\beta=4\pi/3$) are represented by yellow dots in Figs.\,\ref{fig8}(a) and \ref{fig8}(b), respectively. A straight line connecting near-field MIP and center of auto-focused bright spot is drawn, where an arrow denotes the direction of flow of intensity in order to form high-power density lobes. Figures\,\ref{fig8}((c)-(d)) show the intensity distributions of aALB at a distance of $z=70$ cm for ($w=1$ mm, $\beta=7\pi/6$) and ($w=1$ mm and $\beta=4\pi/3$), respectively. The bright lobes in the pattern are marked as 1, 2 and 3. When near-field MIP, auto-focused central bright spot and bright lobe in the pattern are aligned well in a straight line (shown by a solid line in Fig.\,\ref{fig8}(a)), the intensity flows towards that aligned bright lobe in the pattern, and forms a high-power density lobe in that direction. For example, in Fig.\,\ref{fig8}(a) (for $w=1$ mm and $\beta=7\pi/6$), only bright lobe 2 is aligned perfectly, so most of the intensity shifts to only bright lobe 2, and forms a high-power density lobe (Fig.\,\ref{fig8}(c)). In this case, most of the intensity from bright lobes 1 and 2 transfers to bright lobe 2. More details on the flow of intensity and creation of high-power density lobe are shown in the Appendix C. When such alignment does not satisfy, for example, in Fig.\,\ref{fig8}(b) (for $w=1$ mm and $\beta=4\pi/3$), the shifting of most of the intensity in a single bright lobe does not occur, and a significant portion of intensity also remains in the other bright lobes, as shown in Fig.\,\ref{fig8}(d). In this case, the bright bright lobe 2 was more close to the alignment (marked by solid line in Fig.\,\ref{fig8}(b)) as compared to the bright lobe 1, so it receives more intensity. As the arrow direction points from bottom to top, so intensity flows mostly in that direction. In this case, intensity from bright lobe 3 transfers almost completely, and from bright lobe 1 transfers partially.

Note, for each bright lobe in the pattern this alignment can occur for two MIPs (diametrically opposite values in Fig.\,\ref{fig7}), where in one case high-power density lobe is obtained before the auto-focusing distance (at $z=70$ cm), and for the other after auto-focusing point ($z\approx 100$ cm). For example, it has been observed for $\beta = \pi/2$ (Fig.\,\ref{fig3}(h1)) and $\beta = 3\pi/2$ (Fig.\,\ref{fig5}(j1)).
This can be attributed to the fact that there exist an on-axis symmetry around auto-focusing distance ($z_{af}$) for both ideal ALB and aALB (Fig.\,\ref{fig4}(c)). More specifically, the spatial intensity distribution of ideal ALB at $z= 70$ cm (Fig.\,\ref{fig3}(c1)) and $z= 100$ cm (Fig.\,\ref{fig3}(e1)) are similar. However, for aALB, due to intensity migration the region of less intensity in aALB at $z$ = 70 cm  (Fig.\,\ref{fig5}(h1)) becomes a region of high intensity at $z$ = 100 cm (Fig.\,\ref{fig5}(j1)).

In the considered examples of ALB and aALB, there are three bright lobes for $m=3$, which are oriented at different angles. Thus, the high-power density lobes can be created at three different spatial positions by choosing three specific values of $\beta$. For $m = 3$, three $\beta$ values are $(\pi/2 \,, 7\pi/6 \,,  11\pi/6)$ (for distance $z=70$ cm, before the auto-focusing). However, to generate high-power density lobes at more spatial positions, the value of $m$ can be increased further ($m>3$), and accordingly there will be more values of $\beta$.  In general, for a given value of $m$, $\beta$ can be found using the following empirical relation:
\begin{equation}
\mathrm{Odd}\; m \quad \beta_n = \frac{\pi}{2m}(4n - 1) \;\;\; n = 1,2,..m \label{eq21}
\end{equation}
\begin{equation}
\mathrm{Even}\;m \quad \beta_n = \frac{\pi}{2m}(4n - 3) \;\;\; n = 1,2,..m \label{eq22}
\end{equation}

\begin{figure*}[htbp]
\centering
\includegraphics[ width = 13cm, keepaspectratio = true]{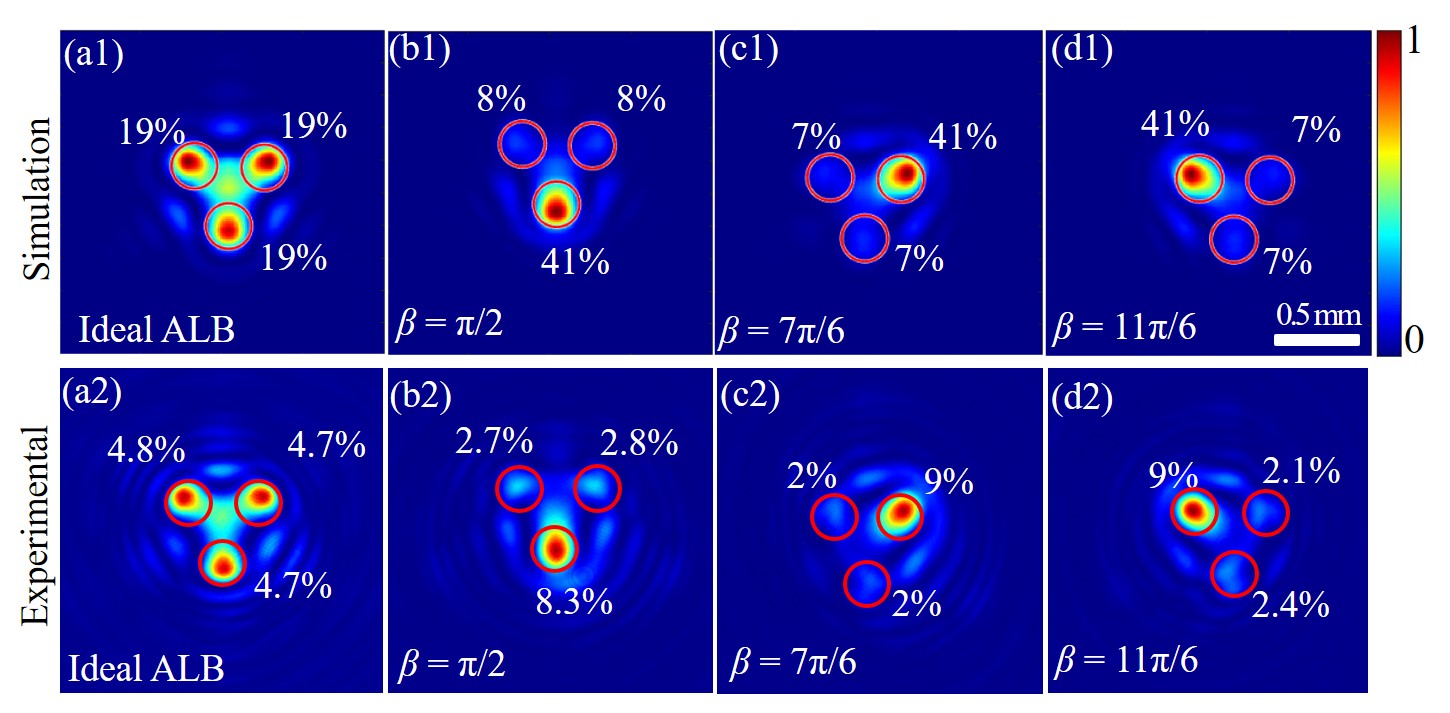}
\caption{(a1, a2) The intensity distributions of ideal ALB at $z=70$ cm. The intensity distributions of aALB at $z=70$ cm for different values of $\beta$ (b1, b2) $\pi/2$, (c1, c2) $7\pi/6$, and (d1, d2) $11\pi/6$. The other parameters: $w=1$ mm, $\alpha = 5.9$ mm$^{-2}$, $\sigma = 1.45$\,mm, $m = 3$, $q = 2$ and $\lambda=632$ nm.}
\label{fig9}
\end{figure*}

Further, for the case of $m=3$, we quantified the shift of intensity in bright lobes for three specific values of $\beta$. The results are shown in Fig.\,\ref{fig9}.
Figures\,\ref{fig9}(a1) and \ref{fig9}(a2) show the simulated and experimental intensity distributions of ideal ALB at $z=70$ cm, respectively, indicating the equal intensity in all three bright lobes. The percentage of intensity inside each lobe (marked by red circle) is calculated by the method of diffraction efficiency \cite{miles2018fabrication}. The diffraction efficiency represents the amount of intensity inside the bright lobe (area marked by red circles) with respect to the total intensity of beam. Note, in the experimental results the intensity inside the bright lobes is distributed equally, but their values are smaller than the numerical results. We attribute this difference due to the imperfections related to SLM, which causes a significant residual reflection. The SLM is anti-reflection (AR) coated at 1064 nm, however, the experimental results are obtained with He-Ne laser at 632 nm, due to the availability in our lab. Figures\,\ref{fig9}((b1)-(d1)) and \ref{fig9}((b2)-(d2)) show the simulation and experimental results of intensity distribution of aALB at $z=70$ cm for three different values of $\beta=\pi/2,~7\pi/6$, and $11\pi/6$, respectively. As evident, for these three specific values of $\beta$, most of the intensity is transferred to any of the bright lobes, spatially positioned at different locations. More specifically, for aALBs, the intensity is enhanced by a factor of $>2$ in any one of these bright lobes (41$\%$) as compared to ideal ALB (19$\%$). Further, asymmetry shifts a major portion of intensity in one of the bright lobe (high-power density lobe), and the intensity between the high-power density lobe (41$\%$) and other lobes ($\sim7\%$) differs by a factor $\sim6$, as shown in Figs.\,\ref{fig9}((b1)-(d1)). The experimental results show qualitatively the same behaviour, however, the difference factor is obtained between $3-4.5$ (Figs.\,\ref{fig9}((b2)-(d2))). 

From these results, it is clear that we can generate high-power density lobes with controlled spatial position by introducing the asymmetry. 

\section{Conclusions}\label{VII}
We have generated asymmetric aberration laser beams with controlled intensity distribution, using a diffractive optical element involving phase asymmetry. More specifically, high-power density lobes are formed with controlled spatial position. It is found that the auto-focusing properties of aALBs remain invariant with respect to the asymmetry parameters ($w$ and $\beta)$. However, the intensity distribution of aALBs strongly depends on these asymmetry parameters. The asymmetry parameters control the position of indeterminate phase point of the trigonometric phase term, which creates a controlled asymmetry in the intensity distribution at the near-field plane. After further propagation it controls the flow of intensity in different parts of the beam. For the specific values of $\beta$, it is possible to transfer and confine most of the intensity in a single bright lobe (high-power density lobe). The spatial position of this high-power density lobe can be controlled with a proper choice of $\beta$. We have determined empirical relations of $\beta$ for even and odd values of $m$, for which a single high-power density lobe can be created at different spatial positions. 

In addition to auto-focusing features, a controlled intensity distribution with high-power density lobes is crucial for many applications. Thus aALBs may be considered suitable for a range of applications viz. guiding of microparticles, material processing , ablation and surgical uses. \cite{Chremmos2011}. 
\appendix
\section{Generalized expression of aALB}\label{appendixA}
The conventional ALB expression \cite{Reddy2020}:
\begin{eqnarray}
U(r,\phi) = \exp(\frac{-r^2}{2\sigma^2})\exp(-i\alpha r^q + i\sin(m\phi)) \; r \leq R. \label{eqA1}
\end{eqnarray}
such that 
\begin{eqnarray}
r\exp(i\phi) = x + iy,~~~ r^2 = x^2 + y^2, \label{eqA2}
\end{eqnarray}
where $q$  represents an arbitrary radial power of dependence, $m$ is a positive integer denotes periodic angular dependence, $\alpha$ is the real parameter, has a dimension of mm$^{-q}$, and $R$ is the radius of the DOE. With the complex Cartesian coordinate shifting:
\begin{equation}
U(x,y) = exp\left(\frac{-(x^2 + y^2)}{2\sigma^2}\right)\exp(-i\alpha s^q + i \sin( m\theta)).\label{eqA3}
\end{equation}
such that 
\begin{eqnarray}
s^2 = (x - x_o)^2 + (y - y_o)^2, \label{eqA4}\\
x_o = a + ib, \quad y_o = c + id. ~~~~ a,b,c,d \in\mathbb{R} \label{eqA5}
\end{eqnarray}
Following the simplifications used in Eqs.(5)-(11), and substituting value of $s^2$ from Eq.\,(9) in the phase part of Eq.\ref{eqA3}, we get
\begin{eqnarray}
\exp(-i\alpha s^q + i\sin(m\theta))\nonumber\\
      =\exp\large\left(-i\alpha(s^2)^{q/2}\right)\exp(i\sin(m\theta)), \nonumber\\
      =\exp(-i\alpha(G\exp(i\gamma)^{q/2})\exp(i\sin(m\theta)), \nonumber\\
      =\exp(-i\alpha G^{q/2}\exp(i\frac{q\gamma}{2})\exp(i\sin(m\theta)),\nonumber\\
      =\exp(-i\alpha G^{q/2} \left(\cos(q\gamma/2)+i\sin(q\gamma/2)\right))\exp(i\sin(m\theta)),\nonumber\\
      =\exp(-i\alpha  G^{q/2}cos(q\gamma/2)+i\sin(m\theta)) \nonumber \\ 
      ~~~~~\times \exp(\alpha G^{q/2} \sin(q\gamma/2)).\label{eqA6}
     \end{eqnarray}
The term $\exp(\alpha G^{q/2}\sin(q\gamma/2))$ is a real quantity, which will not contribute to the complex phase, so we get the following phase expression
\begin{eqnarray}
\exp(-i\alpha s^q + i\sin(m\theta))\nonumber\\
= \exp(-i\alpha\,G^{q/2}\,cos(\frac{q\gamma}{2})+i\sin(m\theta)), \label{eqA7}
\end{eqnarray}
Thus the final expression of aALB becomes
\begin{eqnarray}
\fl U(x,y) &= \exp\left(-\frac{(x^2 + y^2)}{2\sigma^2}\right)\exp(-i\alpha G^{q/2}\cos(\frac{q\gamma}{2})\nonumber \\ 
&+ i\sin(m\theta)). \label{eqA8}
\end{eqnarray}
The values of $G$ and $\gamma$ can be calculated from Eqs.\,(\ref{eq11})-(\ref{eq12}). 

\section{}\label{appendixB}
\begin{figure}[htbp]
\centering
\includegraphics[width = 8.5cm, keepaspectratio = true]{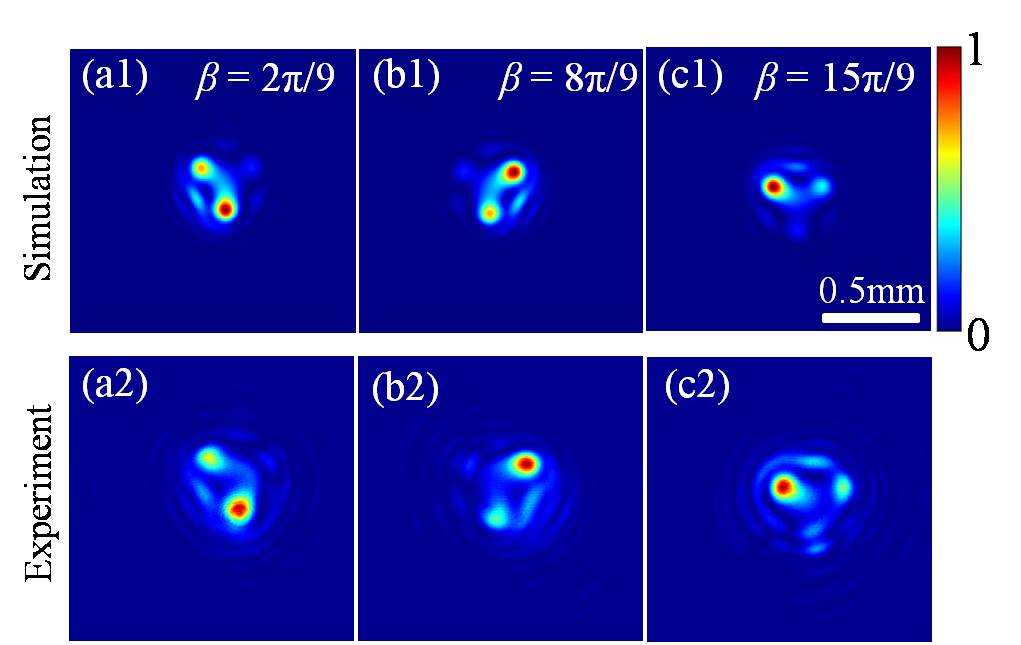}
\caption{The simulation (upper row) and experimental (bottom row) intensity distribution of aALB at $z=70$ cm for $\beta$ (a1, a2) $2\pi/9$, (b1, b2) $8\pi/9$, and (c1, c2) $15\pi/9$. The other parameter values are $w=1$ mm, $\alpha = 5.9$ mm$^{-2}$, $\sigma = 1.45$\,mm, $m = 3$, $q = 2$ and $\lambda=632$ nm.}
\label{figB}
\end{figure}

Figure.\,\ref{figB} shows the simulated ((a1)-(c1)) and experimental ((a2)-(c2)) intensity distributions of aALB at $z=70$ cm for $\beta=2\pi/9,8\pi/9,15\pi/9$. As evident, for different $\beta$ values the intensity distribution in the bright lobes becomes different.

\section{} \label{appendixC}
\begin{figure}[htbp]
\centering
\includegraphics[width = 8.5cm, keepaspectratio = true]{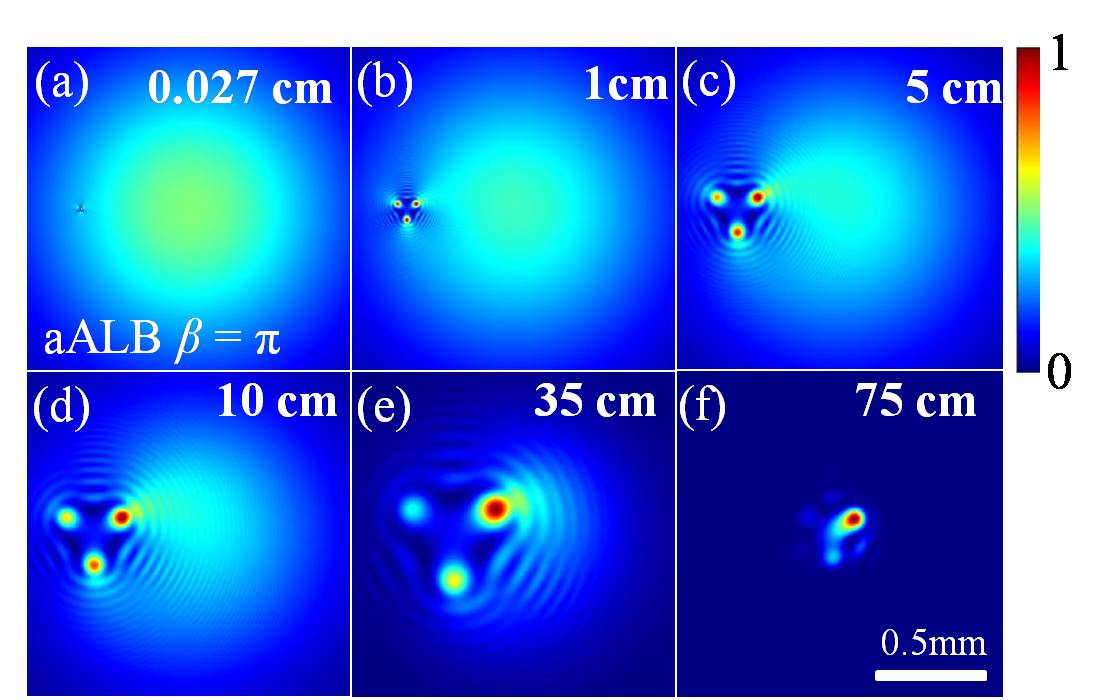}
\caption{Intensity distribution of aALB for $\beta = \pi$ at various propagation distances $z$ (a) $0.027$ cm, (b) $1$ cm, (c) $5$ cm, (d) $10$ cm, (e) $35$ cm, (f) $75$ cm. The other parameter values are $w=1$ mm, $\alpha = 5.9$ mm$^{-2}$, $\sigma = 1.45$\,mm, $m = 3$, $q = 2$ and $\lambda=632$ nm.}
\label{figC}
\end{figure}

Here, we have shown the propagation of aALB from small to large values of $z$. As evident, due to asymmetry a bright peak on the left of Gaussian distribution (MIP) starts to appear (Fig.\,\ref{figC}(a)), which then initiates the development of bright lobes off-centered from the on-axis center (Figs.\,\ref{figC}((b)-(c))). With the propagation the intensity keeps shifting from background to bright lobes. Unlike ideal ALB, in aALB due to asymmetry the intensity shifts inside the bright lobes asymmetrically. For a specific set of asymmetry parameters, it completely shifts in one of the bright lobes. Further, during the propagation over larger distances, the bright lobes shifts toward the on-axis center, as shown in Figs.\,\ref{figC}((d)-(f)).

\section*{References}
\providecommand{\newblock}{}

\end{document}